# Coupled hydro-mechanical aging of short flax fiber reinforced composites

Arnaud Regazzi, Stéphane Corn[*], Patrick Ienny, Anne Bergeret

*École des Mines d'Alès, C2MA, 6 avenue de Clavières, F-30319 Ales Cedex, France*

One of the challenges in the widespread use of biocomposites for engineering applications is the influence of environmental conditions on their mechanical properties, particularly for a combination of aging factors such as temperature, moisture, and mechanical stresses. Thus, the purpose of this paper is to study the influence of coupled aging factors by focusing on a 100% bio-based and biodegradable composites made of flax/poly(lactic acid) with several fiber contents. The development of a specific testing setup enabled continuous in-situ measurements and allowed comparing the effects of combined aging factors to those of uncombined aging factors. It was confirmed that the aging temperature in wet conditions led to a loss of elastic properties, especially for higher fiber fractions. While creep tests in dry conditions resulted in little decrease of elastic properties, it was observed that mechanical loading of the materials combined with water immersion resulted in a strong synergistic effect on the loss of stiffness. Finally, the presence of fibers reduced environmental stress cracking mechanisms and increased the time to failure.

## 1. Introduction

### 1.1. Context

Nowadays, sustainable development is of major concern, and with it, biomaterials have become part of our lives. In the field of composites, a wide range of biopolymers combined with numerous natural fibers allows to meet various criteria in terms of functional properties. Besides using renewable resources, bio-based composites present many advantages such as low environmental impact, low density, and in some cases biodegradable ability [1]. In several industrial applications (i.e. automotive, sports and leisure, construction and infrastructure [2]), these materials are considered as a promising alternative to the usual oil-based materials reinforced with glass fiber [3]. For these reasons they are likely to increase their market share. However, a widespread use of these composites is curbed by technical difficulties [4]. Among them, the most important are fiber cultivation vagaries, manufacture of composites, and imprecise knowledge of their behavior. But most of all, using such materials in real life conditions (e.g. in the prospect of outdoor applications) leads to specific problems of aging due to numerous factors like temperature, water, radiations, bacteria and/or mechanical loadings. These factors are meant to decrease material properties, and to the worst case scenario they may lead to its complete degradation.

Previous works described the various physico-chemical processes that may take place into composites during aging (i.e. plasticizing, swelling, hydrolysis, oxidation, interfacial decohesion) [5]. Although the mechanisms of these processes are properly described, their interactions and their effects on material properties are still poorly known. Consequently the interdependencies between the aging processes render the prediction of mechanical properties in given condition utterly complex. That is why environmental and mechanical aging are mostly studied separately.

Yet some studies focused on the assessment of the coupled influence of water and stress [6–9], mainly on thermosets composites [10–15]. These studies mostly underlined the acceleration of water diffusion in polymer resulting from the applied stress [16]. As a result, repeated mechanical loadings accelerates physico-chemical aging mechanisms and may even induce others [17].

In the literature, studies related to biobased composites durability as a consequence on their mechanical properties have focused mostly on laminate composites made of thermoset matrices [18–23] But thermoplastic composites may also require substantial mechanical properties [24–27]. In this paper, it was proposed to assess the aging of poly(lactic acid), a thermoplastic matrix also named PLA, reinforced with short flax fibers. The main advantages of this composite are its 100% bio-based origin, its biodegradability, its low price, and its mechanical properties equivalent to those of some oil-based polymers (e.g. polyethylene terephthalate (PET)). Several papers have referred to the interesting properties of PLA reinforced with natural fibers [1,28,29].

[*] Corresponding author.
  *E-mail address:* Stephane.Corn@mines-ales.fr (S. Corn).

However, these composites are sensitive to environmental conditions. The thermo-hydric aging of flax/PLA composites was thoroughly discussed in a previous work [30]. Several papers also underline the sensitivity of PLA to damage mechanisms in a hygrothermal environment resulting from its low glass transition temperature [25,31]. Like any polymer, it undergoes plasticizing and swelling during sorption, but temperature may also trigger irreversible mechanisms such as hydrolysis [32]. This process results in polymeric chain scissions and consequently in permanent modifications of its properties [33]. Natural fibers behavior is also decisive in the aging process. Indeed, temperature and humidity hold sway also over the mechanical properties of plant fibers [34,35]. Therefore a wet environment induces mostly a decrease of the mechanical properties of composites reinforced with plant fibers [36,37].

Concerning the influence of long term mechanical loading on the aging of PLA/flax composites, no assessment has been reported yet.

### 1.2. Methodology

The purpose of this paper is to study the coupled thermo-hydro-mechanical aging of flax/PLA composites. It presents firstly the impact on elastic properties of each aging factor considered separately (i.e. water immersion or mechanical loading) and then the impact of the factors concomitantly.

In a first phase of the study, composites with different fiber contents were immersed in order to study the sole influence of water at several temperatures. In a second phase, other samples of the same materials were submitted to creep tests at these same temperatures in order to evaluate the effects of a long term mechanical loading only. Finally, thermo-hydro-mechanical aging was assessed for these materials, and the evolution of their mechanical properties was compared to the direct addition of the separate effects of hydric and mechanical aging.

Mechanical properties were assessed *in situ* in order to simulate service conditions.

## 2. Materials and methods

### 2.1. Materials

#### 2.1.1. Poly(lactic acid)
PLA Ingeo™ 7000D resin was produced by NatureWorks® LLC (Blair, NE, USA). This grade of PLA, designed for injection stretch blow molded applications, had a density of 1.24 g/cm$^3$, a glass transition temperature between 55 and 60 °C and a melting temperature between 155 and 165 °C [38]. Preliminary tensile tests of injected samples showed a Young's modulus of 3.8 ± 0.1 GPa, a strength of 65 ± 1 MPa, and a strain at break of 4.2 ± 0.6%.

#### 2.1.2. Flax fibers
The short flax fibers (*Linum usitatissimum*) FIBRA-S®6A used for this study were provided by Fibres Recherches Développement® (Troyes, France). According to the technical datasheet [39], fiber bundles were 6 mm long with a diameter of 260 ± 150 μm and their density was between 1.4 and 1.5 g/cm$^3$. The Young's modulus of fiber bundles was 36 ± 13 GPa, maximum stress was 750 ± 490 MPa and strain at break was 3.0 ± 1.9%.

### 2.2. Materials and techniques

#### 2.2.1. Processing conditions
Several fiber weight contents were used: 0% (neat PLA) hereafter named PLA, 10% hereafter named PLA-F10, and 30% hereafter named PLA-F30. Polylactic acid granules were dried at 80 °C for 24 h and flax fibers were vacuum dried at 120 °C for 4 h. Composite granules were obtained with a corotative twin-screw extruder (Clextral BC21, screw length = 900 mm; temperature profile along the screw and at the die = 180 °C). After a second drying step under vacuum at 80 °C during 24 h, compounded granules were molded with an injection molding machine (Krauss Maffei KM50-180CX) into dog-bone samples according to the standard ISO 527-2 1BA. The temperature profile was increasing up to 200 °C and the mold was kept at 25 °C. After processing, samples were stored at room temperature and 2%rh (relative humidity) before characterization or aging. This equilibrium state was considered as the reference for evaluating the effects of aging on the materials [30].

#### 2.2.2. Size exclusion chromatography
The molecular mass of PLA was evaluated by size exclusion chromatography (SEC) with Optilab® rEX™ of Wyatt Technology (CIRAD, UMR 1208 (IATE), Université de Montpellier, France). 90 mg of each aged material was diluted in tetrahydrofuran stabilized with butylated hydroxytoluene, and then kept at 30 °C during 40 h in a water bath. After a 0.45 μm-filtration, each solution was injected in the column for measurement. The reproducibility was evaluated on 3 samples.

#### 2.2.3. Mechanical characterization
In order to assess the long term behavior of materials in real use conditions, tensile mechanical loadings were applied. One can consider two types of long-term loading: fatigue [40] and creep tests [41]. The latter was chosen for its ease of interpretation but also to avoid any dynamic effect on the behavior of materials.

The machine used for these tests was a Dartec model 100 kN monitored by a Tema Concept® control system. A specific set-up including a sealed polycarbonate tank was designed to allow these tests to be performed at constant relative humidity as well as underwater (cf. Fig. 1):

i. For atmospheric tests, its top was covered with an elastomeric film to avoid air transfer between the enclosure and the room. In order to regulate relative humidity, the bottom of the tank was filled with silica gels. Temperature was monitored with a thermo-regulated circulator (Julabo CF-31) and water circulated in a flexible hose coiled inside the tank.
ii. For immersed tests, the tank was filled with distilled water and directly connected to a thermo-regulated circulator (Julabo CF-31).

The axial displacement was measured with a linear variable differential transformer (LVDT) fixed on the upper clamp above the tank. Its core was attached to a rod linked to the lower clamp. The measurements resolution were 1 μm for the LVDT and 0.1 N for the 50 daN force sensor.

Given that the ultimate stress of PLA was around 65 MPa (cf. 2.1.1), the creep tests were carried out at a constant stress of 10 MPa during 144 h or until failure of the sample. Their stiffness was monitored during this aging by performing partial unloading-reloading cycles on the samples every 30 min (ramps being set at 50 N/s). The elastic modulus was determined during unloading

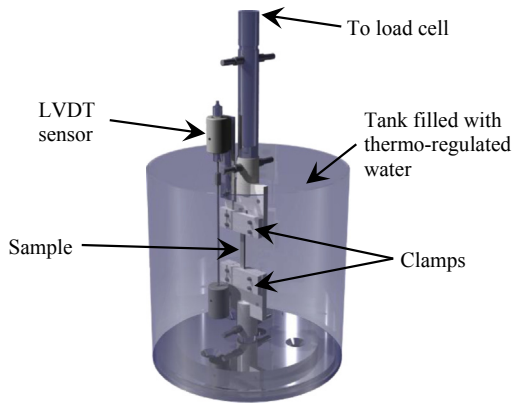

**Fig. 1.** The set-up used for environmental creep tests.

from the slope of the stress-strain curve between 10 and 5 MPa.

The duration of the aging tests led to limit the assessment of the reproducibility of the measurements to only one set of aging conditions (i.e. for PLA-F10 in immersion at 20 °C). The total strain and the elastic modulus of two samples of flax/PLA composites during hydro-mechanical aging are presented in Fig. 2. Creep strain and elastic modulus turned out to be quite reproducible, the relative deviation being less than 10% in term of total strain and 2% in term of elastic modulus. Finally, the lifetime of samples differed only from 6%.

## 3. Results

### 3.1. Thermo-hydric aging

Prior to any assessment of the coupled thermo-hydro-mechanical aging, it was necessary to understand the sole influence of water on elastic properties in relation with aging temperature. The set-up described in 2.2.3 was used in its hydrothermal configuration. The morphology of PLA was assessed during aging by evaluating its number average molar mass by size exclusion chromatography [30]. Fig. 3 shows the number average molar mass of PLA for each material and at each aging temperature before and after 24 h and 144 h of aging.

The initial molar mass of materials slightly decreased with the fiber content because of the natural presence of water in flax fibers despite drying, resulting in a more significant hydrolysis during processing. During aging, the loss of PLA molar mass caused by hydrolysis was globally limited and concerned only composites. For PLA-F10, the decrease only occurred at 50 °C. For PLA-F30, a significant decrease was observed for all temperatures during the first 24 h, but after this period the drop stopped at 20 and 35 °C.

Fig. 4 displays the evolution of the *in situ* elastic modulus during immersion in water for each material and each aging temperature. Concerning initial moduli, as expected, results showed a significant improvement with fiber content (+17% and +97% for an addition of 10 and 30%wt, respectively). During thermo-hydric aging, the modulus of every material decreased. This decrease was more significant as the fiber content increased. But the bath temperature turned out to be the most decisive factor responsible for the drop of modulus for all materials. At 50 °C, rigidity became negligible for all materials after only a few hours. Due to this drastic change of behavior, this temperature was not suitable for thermo-hydro-mechanical aging since any test would end up after only few minutes. Consequently, mechanical aging was only conducted at 20 and 35 °C.

### 3.2. Thermo-mechanical aging

In this section, the study is focused on the influence of the sole mechanical loadings. The set-up described in 2.2.3 was used in the atmospheric configuration for the following discussion. Fig. 5 presents the total strain of composites during creep tests at 10 MPa in dry air at 20 and 35 °C. For each test, the stress-strain response exhibited a quick primary creep (corresponding to the decrease of the creep rate) and an extended secondary creep (where the creep rate remains constant). Neither break nor tertiary creep (corresponding to an increase of the creep rate) was observed on any samples.

Strain rate was influenced by both temperature and fiber content. Indeed, a lower fiber content resulted in a higher strain rate. However, increasing the temperature resulted in a more significant increase of the strain rate. Regarding the impact on the elastic modulus, Fig. 6 shows that no change occurred for all materials at 20 °C. However, at 35 °C, it decreased slightly and even more as fiber content was increased.

As a result, based on the absence of both tertiary creep and modulus change at 20 °C, one can assume that no damage was induced on materials when applying a 10 MPa stress at this temperature. However, at 35 °C, the damage on elastic behavior was real but still limited.

### 3.3. Thermo-hydro-mechanical aging

After studying the influence of thermo-hydric and thermo-mechanical aging separately, it was finally possible to apprehend the coupled thermo-hydro-mechanical aging of flax/PLA composites. The set-up described in 2.2.3 was used in its immersed configuration.

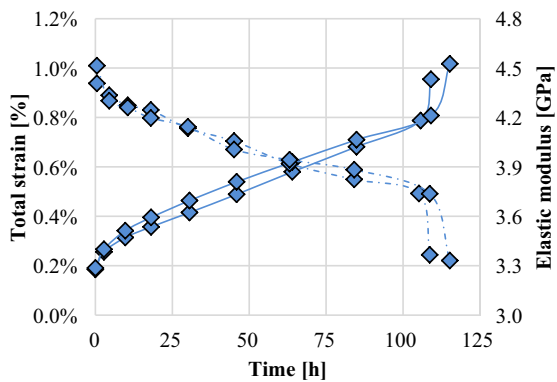

**Fig. 2.** Evaluation of the reproducibility of the total strain (continuous lines) and the *in situ* elastic modulus (dash-dotted lines) for PLA-F10 during creep tests at 10 MPa in water.

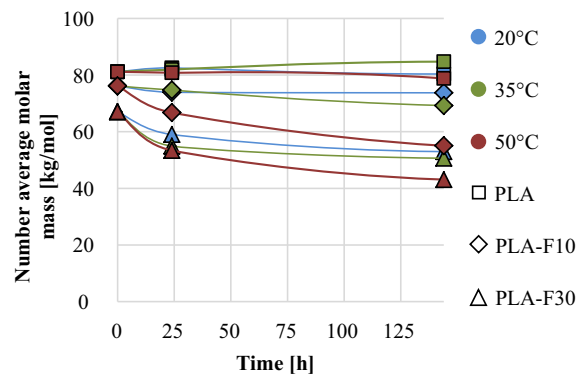

**Fig. 3.** Number average molar mass of PLA in flax/PLA composites depending on immersion time in water and temperature.

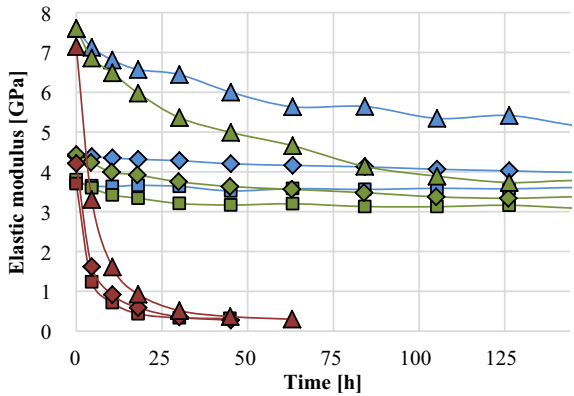

**Fig. 4.** In situ elastic modulus of flax/PLA composites depending on immersion time in water and temperature.

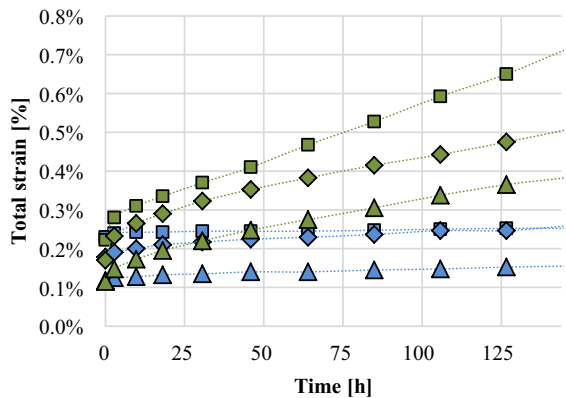

**Fig. 5.** Total strain of composites flax/PLA during creep tests at 10 MPa in dry air (10% rh); the legend is the same as in Fig. 3.

### 3.3.1. Strains

Fig. 7 highlights the differences of the creep responses at 20 °C measured in the presence or the absence of water (i.e. the difference between mechanical and hydro-mechanical aging). For all materials, the presence of water resulted in a drastic increase of the strain rate as a result of plasticizing.

Besides, the lifetime of materials subjected to coupled aging turned out to be shortened compared to tests carried out in dry air. As shown in Table 1, virgin PLA broke after about one day of immersion, although diffusion was still incomplete [30]. PLA-F10 withstood barely 5 days of hydro-mechanical aging before failure. However, no failure was recorded for PLA-F30 before ending the experiments. Consequently, lifetime was strongly dependent of the fiber content and it increased drastically with fiber content. However, no clear trend could be established regarding the influence of the fiber content on strain rate. Neat PLA was the only material which exhibited tertiary creep before failure. A visual inspection of the samples showed by transparency the occurrence of crazes oriented transversely to the loading direction (c.f. Fig. 8). Their distribution seemed homogeneous along the length of the samples, but, during tertiary creep, their concentration increased in the vicinity of the failure area.

At 35 °C, the observations were different. Fig. 9 shows the creep response of samples immersed in water at 35 °C. A drastic increase of the strain rate was exhibited for samples which underwent hydro-mechanical aging. The total strain exceeded 10%. During all creep steps, lower fiber contents exhibited faster strain rates. Secondary creep was very short, even nonexistent, and gave way to an extensive and strong tertiary creep. As shown in Table 2, the lifetime of samples was increased compared to samples immersed at 20 °C. Indeed, PLA only failed after 3 days of thermo-hydro-mechanical aging which was about 3 times longer than PLA in water at 20 °C. Finally, composites aged at 35 °C did not break in the duration of the experiments.

### 3.3.2. Elastic modulus

The evolution of elastic modulus was also assessed during these experiments. In the first place, the case of the PLA-F10 at 20 °C can be taken as an example. Fig. 10 shows separately the influence of the creep stress, of the presence of water, and of the coupling of both these aging factors on the evolution of the elastic modulus of PLA-F10 at 20 °C. As mentioned in 3.2, the influence of the sole creep stress was minor whereas water had a much larger impact on elastic modulus. When aging factors were coupled, the modulus loss was obviously greater than the losses induced by each aging factor taken separately but also greater than their addition. The difference between this addition and the actual loss was the evidence of what we call a synergistic degradation.

In the purpose of an easier understanding, only modulus losses are reported in Tables 1 and 2. They were evaluated at failure, or at the end of the experiment (144 h) if failure did not occur. Globally, the modulus drop increased with both fiber content and temperature. Finally, the synergistic degradation seemed to be relatively independent of the aging temperature. However, it decreased significantly with fiber content.

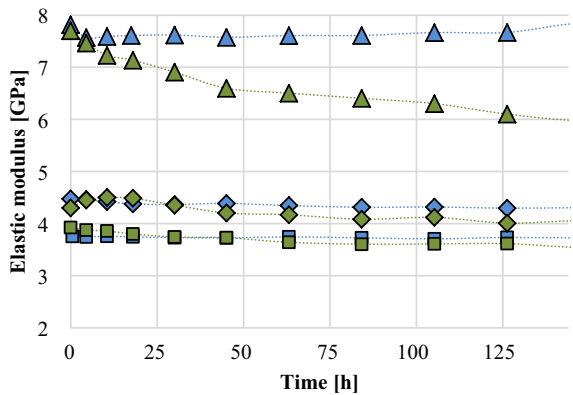

**Fig. 6.** In situ elastic modulus of composites flax/PLA during creep tests at 10 MPa in dry air (10%rh); the legend is the same as in Fig. 3.

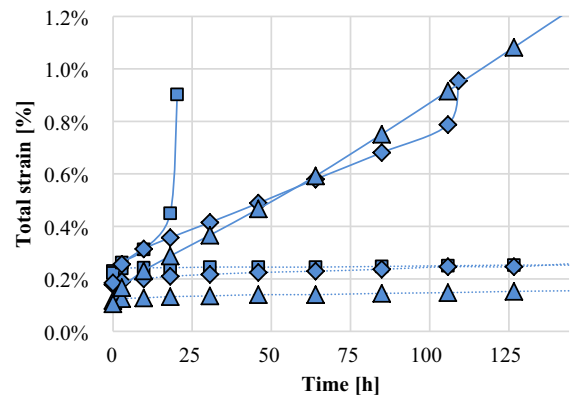

**Fig. 7.** Total strain of composites flax/PLA during creep tests at 10 MPa and 20 °C, in dry air (dotted lines) and in water (continuous lines); the legend is the same as in Fig. 3.

**Table 1**
In situ modulus loss and time to failure of composites flax/PLA caused by the different aging factors during thermo-hydro-mechanical aging at 20 °C.

| Material | Creep | Immersion | Coupled creep and immersion | Synergistic degradation | Failure |
|---|---|---|---|---|---|
| PLA | −1.3% | −5.0% | −27.4% | −21.1% | 20 h |
| PLA-F10 | −3.8% | −10.7% | −25.5% | −11.0% | 115 h |
| PLA-F30 | −1.3% | −31.9% | −38.3% | −5.1% | none |

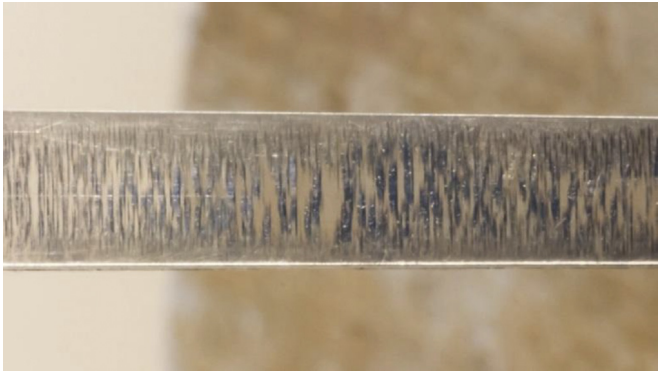

**Fig. 8.** Apparition of crazes into a sample of PLA exposed to a creep stress of 10 MPa in immersion at 20 °C.

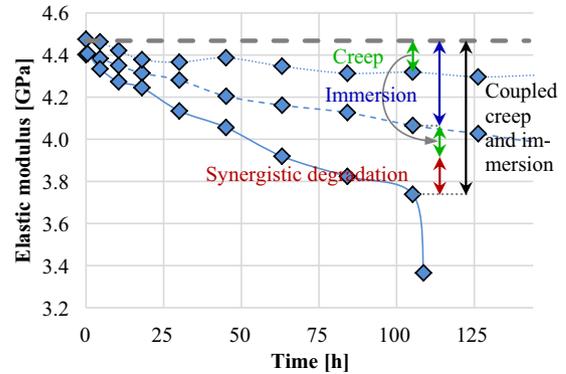

**Fig. 10.** In situ elastic modulus of PLA-F10 at 20 °C depending on immersion time in water (dashed line), and during creep tests at 10 MPa in dry air (dotted line) and in water (continuous line).

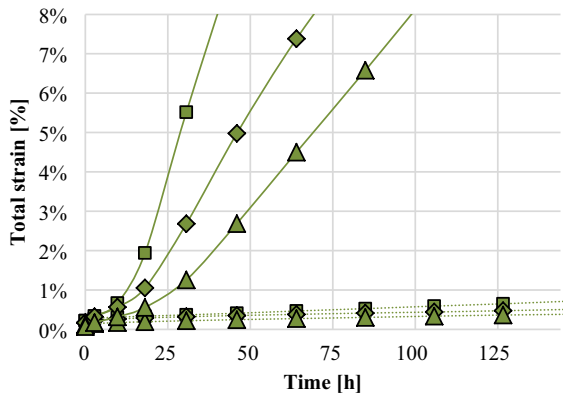

**Fig. 9.** Total strain of composites flax/PLA during creep tests at 10 MPa and 35 °C, in dry air (dotted lines) and in water (continuous lines); the legend is the same as in Fig. 3.

## 4. Discussion

It has been shown that the evolution of the elastic modulus during thermo-hydric aging was the consequence of several simultaneous phenomena which were reversible or irreversible [30]. At 20 °C, the only occurring phenomenon was plasticizing due to the physical presence of water. Consequently, the limited evolution of the PLA modulus at low aging temperature indicated that matrix plasticizing was relatively limited. However, the significant plasticizing of fibers suggested in the literature [3,42,43] explained the decrease of the elastic modulus with the fiber content. When increasing aging temperature, both reversible and irreversible effects grew. At 50 °C, the proximity of the PLA glass transition temperature combined with plasticizing (which lowers the temperature of this transition [44]) were both responsible for a loss of interaction between macromolecular chains making materials malleable and viscous. The proximity of the PLA glass transition temperature was also responsible for the hydrolysis of PLA as shown by the decrease of its number average molar mass.

Several results lend credence to the hypothesis that the creep behavior of composites was mostly influenced by the matrix behavior. Firstly, the change in strain rate at 20 °C was found to be rather limited when varying the fiber content. The significant acceleration of the strain rate at 35 °C corroborated this assumption on account of the proximity of the PLA glass transition temperature. Besides, the strain rate of hemp fibers turns out to be particularly low [35]. Consequently, assuming the creep behavior of flax and hemp fibers is similar, it could explain why the strain rate of composites decreased with the fiber content (particularly at 35 °C).

In the meantime, the results in Fig. 6 showed that the decrease of the elastic modulus was independent from the strain rate. Indeed, the higher strain rate of PLA at 35 °C compared to its composites was not concomitant to a more significant decrease of its elastic modulus. Since most of the decrease of the elastic modulus of composites cannot be attributed to the behavior of PLA, it originated either from the behavior of fibers themselves or from the degradation of their interface with PLA. But at 10 MPa (far below the tensile strength), loading and unloading of flax fibers did not induce damage regarding the elastic modulus. On the contrary, fibrils reorientations tend to result in an increase of the elastic modulus [45]. As a result, the damage responsible for the decrease of elastic modulus was most probably occurring at the fiber/matrix interface.

**Table 2**
In situ modulus loss and time to failure of composites flax/PLA caused by the different aging factors during thermo-hydro-mechanical aging at 35 °C.

| Material | Creep | Immersion | Coupled creep and immersion | Synergistic degradation | Failure |
|---|---|---|---|---|---|
| PLA | −4.2% | −17.1% | −43.4% | −22.1% | 75 h |
| PLA-F10 | −10.1% | −25.1% | −47.7% | −12.5% | none |
| PLA-F30 | −18.0% | −50.7% | −68.8% | −0.1% | none |

During thermo-hydro-mechanical aging, the creep behavior in water was very different from the one in air. The increase of strain rate was attributed to the plasticizing of both matrix and fibers. However, the sudden growth of the crazes amount in the polymer matrix during tertiary creep was probably the reason of the acceleration of the strain rate. The high concentration of crazes in the vicinity of the fracture is supposed to be responsible for the brittle failure of samples during this coupled aging. Nevertheless, the presence of fibers prevented an important local concentration of crazes, thus limiting the amplitude of tertiary creep and especially extending the lifetime of samples. Besides, results showed that increasing aging temperature allowed extending lifetime during hydro-mechanical aging. It is most likely that this phenomenon was linked to the plasticizing of composite components by both water and temperature. Its main effect was a more ductile behavior of the material, making it more accommodating to large deformations.

The possible increase of the strain rate with the fibre content in water at 20 °C observed in Fig. 7 could be attributed to the greater sensitivity of the flax fibre modulus to the presence of water at this temperature [42]. This could also result from the lower PLA molar mass in PLA-F30 samples compared to PLA-F10 samples as shown in Fig. 3.

Finally, the synergistic degradation reported in Tables 1 and 2 was attributed to environmental stress cracking for several reasons [46]. Firstly, the appearance of crazes on the polymer was typical of this phenomenon. Secondly, the fragile failure of materials was also characteristic, and excluded any process caused by plasticizing. Then, this phenomenon usually affects amorphous polymers. Indeed PLA was mostly amorphous at this stage of aging [30]. Finally, the highly probable orientation of polymer chains in the direction of the stress due to the sample forming process was consistent with the mechanisms of environmental stress cracking [47].

Regardless to the prevailing mechanism responsible for the synergistic degradation, it was considerably limited by the presence of fibers, probably because they postponed crack propagation into the matrix. Indeed, neat PLA turned out to be very sensitive to this phenomenon, and even more when temperature was decreased. Its fragile behavior at low temperature contributed to this phenomenon, while it became less sensitive close to the glass transition due to the raise of its ductility.

## 5. Conclusions

This study examined the influence of thermo-hydro-mechanical aging on elastic and creep behaviors of poly(lactic acid) (PLA)/flax fibers composites.

Firstly, the sole influence of water was assessed depending on temperature. This thermo-hydric aging turned out to have a major impact on the mechanical properties. The occurrence of both reversible and irreversible phenomena explained the loss of rigidity of these composites. In this case, the elastic behavior was driven by the sensitivity of flax fibers to this type of aging. However, close to its glass transition temperature (about 56 °C), PLA lost its rigidity, making it useless in these conditions for any structural application.

Then the sole influence of a constant stress was evaluated depending on temperature. A creep stress of 10 MPa induced limited damage on all materials in a dry environment, especially compared to thermo-hydric aging. However, the little damage observed mostly on composites was attributed to fiber/matrix interface degradation.

Finally, the coupling of these aging factors was assessed. In this case, hydro-mechanical aging induced a synergy attributed to environmental stress cracking. This effect was characterized by a decrease of the elastic modulus which was more significant than the addition of the combined effects of the two aging factors taken separately. Both temperature and fiber content tended to induce a more significant decrease of the elastic modulus, yet they resulted in extended time to failure. The increase in ductility caused by temperature and the limitation of crack propagation resulting from the presence of fibers were responsible for these observations. As a result, fibers played a predominant role in the durability of composites by limiting environmental stress cracking.